\documentclass[
 aps, pra,
 amsmath,amssymb,
 11pt,
 final,
tightenlines,
 twoside,
 onecolumn,
 nofloats,
nofootinbib,
 superscriptaddress,
showkeys,
showkeywords,
 ]
{revtex4-2}

\usepackage[T2A]{fontenc}
\usepackage[utf8]{inputenc}
\usepackage[russian,english]{babel}
\usepackage{graphicx}% Include figure files
\usepackage{dcolumn}% Align table columns on decimal point
\usepackage{bm}% bold math

\usepackage{ulem}

\input{maik.rty}

\setcitestyle{authoryear,round}
\def\squareforqed{\hbox{\rlap{$\sqcap$}$\sqcup$}}

\def\sq{\ifmmode\squareforqed\else{\unskip\nobreak\hfil
\penalty50\hskip1em\null\nobreak\hfil\squareforqed
\parfillskip=0pt\finalhyphendemerits=0\endgraf}\fi}

\def\utw{\smash{\rlap{\lower5pt\hbox{$\sim$}}}}

\def\udtw{\smash{\rlap{\lower6pt\hbox{$\approx$}}}}

\def\diameter{{\ifmmode\mathchoice
{\ooalign{\hfil\hbox{$\displaystyle/$}\hfil\crcr
{\hbox{$\displaystyle\mathchar"20D$}}}}
{\ooalign{\hfil\hbox{$\textstyle/$}\hfil\crcr
{\hbox{$\textstyle\mathchar"20D$}}}}
{\ooalign{\hfil\hbox{$\scriptstyle/$}\hfil\crcr
{\hbox{$\scriptstyle\mathchar"20D$}}}}
{\ooalign{\hfil\hbox{$\scriptscriptstyle/$}\hfil\crcr
{\hbox{$\scriptscriptstyle\mathchar"20D$}}}}
\else{\ooalign{\hfil/\hfil\crcr\mathhexbox20D}}%
\fi}}

\def\be{\begin{equation}}
\def\ee{\end{equation}}
\def\ba{\begin{eqnarray}}
\def\ea{\end{eqnarray}}

\def\msun{M_\odot}

\def\ltsima{$\; \buildrel < \over \sim \;$}
\def\simlt{\lower.5ex\hbox{\ltsima}}
\def\gtsima{$\; \buildrel > \over \sim \;$}
\def\simgt{\lower.5ex\hbox{\gtsima}}

\usepackage[dvipsnames]{color}
\begin{document}

\selectlanguage{english}

%\keywords{shock waves -- supernova remnants -- dust}

%\ydk{}
%\titlerunning{}
%\authorrunning{}
%\toctitle{}
%\tocauthor{}

\title{High velocity dust grains produced by a supernova explosion}

\author{\firstname{E.~O.}~\surname{Vasiliev}}
 \email{eugstar@mail.ru}
 \affiliation{Lebedev Physical Institute, Russian Academy of Sciences, Moscow, 117997 Russia}

\begin{abstract}
Motivated by the discovery of a meteor with velocity more than 300~km/s \citep{Afanasiev2007} we study how dust grains formed in the supernova (SN) ejecta at its free-expansion phase can keep their high velocity obtained from the gas, where these grains were born. Using a one-dimensional spherically symmetric Lagrangian approach we follow the dynamics of grains during the interaction of a SN ejecta with the ambient interstellar gas. We found that grains larger $\sim 0.3~\mu$m can cross the SN forward shock front and reach the ambient interstellar medium. They are destroyed inefficiently, thus, their sizes remain close to the initial ones. In a diffuse interstellar gas of density $\simlt 1$~cm$^{-3}$ grains larger as $\simgt 1~\mu$m move with velocity around $1-2$ thousand km/s and spread over distances exceeding $\sim 40-90$~pc from the SN origin within tens of kiloyears. In dense environment the propagation of grains is significantly suppressed: even large grains do not spread over distances greater than 50\% of the current SN remnant size. Grains born close to the edge of the SN ejecta core in less-massive ejecta reach greater distances. Thus, high-velocity grains are expected to be formed in this part of the ejecta with mass lower than $\sim 3~\msun$ of the SN remnant evolving in a diffuse gas with density $\simlt 1$~cm$^{-3}$. We discuss possible consequences for dust transport out from galactic disks.
\end{abstract}

\maketitle

%----------------------- Section 1 -------------------------------
\section{INTRODUCTION}\label{sec:intro}

Interstellar dust is produced in significant quantities in ejecta of Type II supernovae  \citep{Kozasa1989,Kozasa2009,Todini2001,Nozawa2003,Nozawa2006,Nozawa2007,Sarangi2015,Sluder2018}. Inside a supernova (SN) remnant grains are formed at early ages \citep{Cherchneff2009,Cherchneff2010,Cherchneff2011,Cherchneff2013,Dwek2016} and desroyed by sputtering during further time \citep{Scalo1977,Shull1977,Tielens1987,Dwek1996,Jones1996,Janev2001}. The dust destruction efficiency  depends significantly on the passage of grains through the reverse shock \citep{Nath2008,Silvia2010,Silvia2012,Biscaro2016,Kirchschlager2019,Kirchschlager2020,Shchekinov2025}. The dust transport to the interstellar medium (ISM) is determined by the SNR evolution \citep{Nozawa2007,Bocchio2016,Micelotta2016,Slavin2017,Slavin2020,vs2024}. 

The ejecta of core-collapse SNe are clumpy \citep{Stathakis1991,Spyromilio1994,Matheson2000}. The velocity range for the oxygen line emission extends to $1500$~km~s$^{-1}$ in SN 1987A and $\sim 4000$~km~s$^{-1}$ in SN 1993J. The oxygen-rich SNRs like Cas~A show evidence for freely expanding, oxygen-rich ejecta in clumps \citep{Chevalier1979}. High-velocity clumps have also high density around $\sim 20-1000$~cm$^{-3}$ and contain a significant amount of dust \citep{Sutherland1995,Docenko2010,Silvia2010,Silvia2012,Biscaro2014,Biscaro2016,Micelotta2016}. The observed clump radii are in the range $(0.5-2.5) \times 10^{16}$~cm \citep{Fesen2011}. These clumps can move like bullets inside the remnant and even ahead the forward shock \citep[e.g.,][]{Kundt1988,Anderson1994,Wang2002}. During this movement they can be stripped due the Kelvin-Helmholtz instability. Dust inside these flows can be destroyed in hot interclump medium \citep{Silvia2010,Silvia2012,Kirchschlager2019}. Large grains are even able to overcome the forward shock due to their inertia \citep{Slavin2004,Slavin2020} and then probably travel into the ISM on large distance from the SN origin. 

It is possible that such grains can transport several isotopes in the Solar system. Studies of the possible biological effects of a near-Earth SN have a long history \citep[e.g.,][]{Shklovskij1969,Alvarez1980,Ellis1995,Melott2017,Fields2019}. One of them is a possible delivery of long-lived radioactive isotopes (with the decay time about several Myr) such as $^{60}$Fe and $^{244}$Pu from SN onto Earth \citep[e.g.,][]{Ellis1996,Breitschwerdt2012,Breitschwerdt2016,Feige2018,Schulreich2017,Schulreich2018}. 
More recently, \citet{Fry2020} have considered the transport of dust grains containing $^{60}$Fe from a SN to distances around 100~pc. They inject grains into the environment of an SNR and trace their trajectories inside the remnant by applying a 1D hydrodynamic description and then through the magnetized ISM using a simplified magnetic field description. They found that the presence of a magnetic field within the shocked ISM material limits the passage of SN dust grains. 

During the scheduled observations the spectrum of a meteor entering to the Earth atmosphere with velocity more than 300~km/s has been firstly detected \citep{Afanasiev2007}. This value is several times higher than the mean geocentric velocity of meteors \citep{Galligan2005}, while it is comparable with the velocity of the Solar system to the Local group \citep{Karachentsev1996}. Based on this \citet{Afanasiev2007} concluded that the grain comes from the intergalactic medium. An estimate of the grain size gives too large value, $a\sim 10^{-2}$~cm, compared to the typical size of the interstellar dust particles. Thus, this detection arises questions about how so massive grains can be formed and how they get such high velocity.

According to the classic nucleation theory, grains of various composition are formed during first several hundred days with a lognormal distribution in a quite wide size range: $\sim 0.001 - 1~\mu$m  \citep{Kozasa1989,Todini2001,Nozawa2003,Bianchi2007,Nozawa2006}. During the condensation of Si and Fe in the inner ejecta, most large grains are formed up to $1~\mu$m \citep[e.g.,][]{Nozawa2003} and even up several microns \citep{Gall2014}. There is a suggestion about the production of large grains in a possible ejection of red giant's hydrogen shell during the collapse of failed supernova \citep{Kochanek2014}.
 
Dust grains with a size of $\sim 30~\mu$m and hyperbolic velocities relative to the Sun were firstly detected by radars as they passed through the Earth's atmosphere \citep{Taylor1996,Baggaley2000,Baggaley2004}. Indications of the existence of large dust grains ($\sim 0.5-1.5~\mu$m) also follow from the observations on board the Ulysses and Galileo sattelites \citep{Landgraf2000,Draine2009rev,Krueger2015,Baalmann2025} and a possible interpretation of some measurements of the infrared background \citep{Wang2015}. As a result the mass distribution of large dust grains can be extended substantially beyond the standard value $0.1~\mu$m \citep[e.g.,][]{Murray2004,Draine2009}. The possibility of the detecting larger interstellar dust grains with sizes up to $\sim 1$~mm was discussed in \citep{Murray2004,Socrates2009}.

In general, supernovae can accelerate interstellar dust to subrelativistic speeds $\simlt 0.1c$ \citep{Spitzer1949, Wolfe1950,Bingham1999,Weiler2003,Hoang2015,Cristofari2024}. Grains located to the SN origin closer as $10^{16}$~cm get high velocities due to the extreme radiation pressure from the SN explosion \citep{Spitzer1949}. However, first of all, dust grains located so close to the SN origin are believed to be swept out by the radiation from massive star -- a SN progenitor. In any case, such grains could be useful for studying nearby SN \citep[e.g.][]{Knie2004,Thomas2016,Melott2017} and constraining SN models \citep[e.g.][]{Wesson2015,Bevan2017,DeLooze2017,Kirchschlager2019}. They could affect on the lunar surface \citep{Siraj2020a} and appear as meteors in the Earth’s atmosphere \citep{Siraj2020b}. 

More usual mechanisms of grain acceleration are connected with strong shocks \citep{Epstein1980,Ellison1997}. As a result the velocity of grains can reach several thousands km/s. Note that grain destruction following the acceleration could play a role in the production of cosmic rays \citep{Cesarsky1981,Bibring1981}. In a magnetized turbulent ISM, grains can be essentially accelerated by various processes \citep[e.g.][]{Lazarian2002,Yan2004,Shukla2003,Prudskikh2008,Kostyukova2010,Mattsson2022,Moseley2023}.

While more naturally, high velocity of grains is related to their place of birth. As it is mentioned above, grains are produced in SN ejecta, when it expands freely with velocity of several thousands km/s. Thus, grains formed in the ejecta have the same velocity and remain coupled with high-velocity ejecta for some period. Afterwards, they interact with the reverse shock and propagate inside hot environment of a SN remnant. There is a question whether grains keep such high velocity during their passage through the remnant and further. Motivated by the detection of a meteor with velocity more than 300~km/s we study the dynamics of grains formed in SN ejecta from their birth to the age of several thousand years, i.e. when SNR becomes radiative. Section~2 describes the model and initial conditions. Section~3 presents the results. Section~4 discusses the application of the results and their consequences. Section~5 summarizes the main conclusions.

%----------------------- Section 2 -------------------------------
\section{MODEL DESCRIPTION}\label{sec:mod}

We use a one-dimensional spherically symmetric Lagrangian approach to follow the evolution of a SN ejecta interaction with the ambient interstellar gas from early time, when the ejecta expands freely and dust begins to be produced, $t_i \sim 100$~days \citep{Kozasa1989}, to the end of the adiabatic phase, i.e. to timescale of the order of the cooling time $t_c \sim k_B T / n\Lambda$, where $k_B$ is the Boltzmann constant, $T$ and $n$ are the temperature and density of a gas behind SN forward shock. For the solar metallicity the cooling time is equal to $t_c \sim 3\times 10^4 n^{-1}$~yr for $T\sim 10^6$~K and $\Lambda \sim 10^{-22}$~erg~s$^{-1}$~cm$^3$. 

We consider the evolution of the SN ejecta with initial kinetic energy $E_k = 10^{51}$~erg, total mass $M_e = 1, \ 3$ and $10~\msun$. Thus, the maximum expansion velocity is given by $v_e = \sqrt{E_k/M_e}$. The density of the ambient gas is set to 1 and 10~cm$^{-3}$, so the final time of the numerical runs is 30 and 3~kyr, respectively. The ambient gas temperature is equal to $10^4$~K. Note that the choice of values for the ejecta masses is determined by both theoretical models \citep[e.g.,][]{Rauscher2002} and interpretation of the observations \citep[e.g.,][]{Lee2014}, with the latter being associated with significant uncertainty. For instance. in the study of dust dynamics in the SN remnant \citep{Slavin2020} the ejecta mass was $3.5\msun$.

\begin{figure*}
\center
\includegraphics[width=10.5cm]{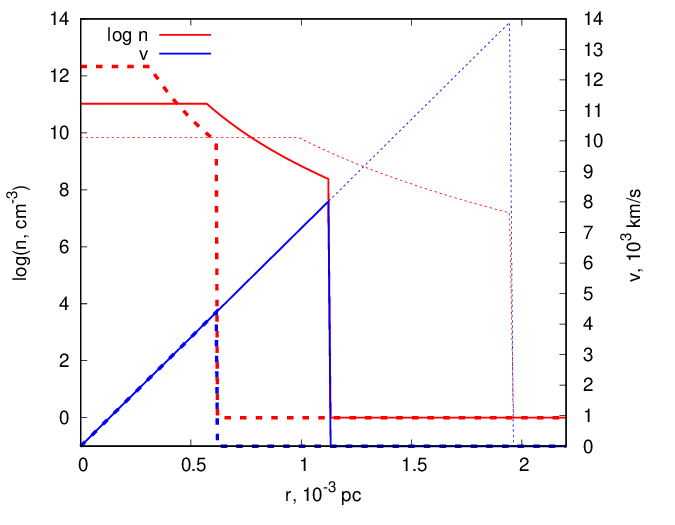}
\caption{
The gas density (left axis) and velocity (right axis) radial self-similar profiles at $t=100$~days for the SN ejecta mass of $M_{e} = 1, \ 3$, and 10~$\msun$ (dot, solid and dash lines, respectively).
}
\label{fig-ini-pro}
\end{figure*}

For the initial distribution of gas density and velocity we adopt the radial profiles following from a self-similar solution for free expansion of the SN ejecta \citep{Truelove1999,ZeldRai}:
\be
v(r) = \left\{ 
                \begin{aligned}  
                   {r \over R_e}v_e, r<R_e \\
                   0  , r>R_e \\
                \end{aligned} 
         \right.
\label{eq-vel}
\ee
\be
\rho(r) = \left\{ 
                \begin{aligned}  
                   {M_e \over R_e^3} f(r/R_e), r<R_e \\
                   \rho_0  , r>R_e \\
                \end{aligned} 
         \right.
\label{eq-rad}
\ee
where $\rho_0$ is the background density, $R_e = v_e t_i$ is the ejecta radius, $f(r/R_e)$ is an arbitrary function, which is taken as
\be
f(r/R_e) = A (r/R_e)^{-\alpha}, r>r_c 
\ee
where $r_c$ is the core radius, we adopt $r_c = 0.5R_e$, $A$ is the normalization 
\be
A = {3 \over 4\pi}\left({1-\alpha/3 \over 1-(r_c/R_e)^{3-\alpha}}\right)
\ee
We adopt the index $\alpha=9$. The gas pressure is set to be negligible \citep{Truelove1999}, for the definiteness we adopt the temperature equal to $T(r) = 10^4$~K. Figure~\ref{fig-ini-pro} presents the initial density and velocity profiles for several masses of the ejecta. In the density profile there are three regions: the core with the flat density distribution, the ejecta envelope with steep power-law decrease of density and the ambient gas. The two former expand freely with radially increased velocity. The outer medium is at rest.

%%%%%%%%%%%%%%%%%%%%%%%%%%%%%%%%%%%%%%%%%%%%%%%%%%%%%%%%%%%%%%%%%%%%%
%%%%%%%%%%%%%%%%%%%%%%%%%%%%%%%%%%%%%%%%%%%%%%%%%%%%%%%%%%%%%%%%%%%%%

We solve a 1D gas dynamical equations using the standard second-order accurate Lagrangian finite-difference scheme of constant mass zones \citep{Bowers1991,Thoul1995,Truelove1999}. We consider two regions: the SN ejecta and the ambient medium. The number of zones (shells) in the ejecta is the same for all runs and equal 110. In the ambient medium the number of shells is changed for different models. To follow the evolution of SN ejecta with age from less than one year to thousands years we set different spatial resolution in these two regions. Initially the sizes of the zones differs about a factor of several hundreds. Thus, the total number of zones in our runs remains reasonable and varies from 2000 to 10000. We consider the grids with a factor of several to tens to investigate a possible dependence on spatial resolution and find a good convergence for a factor of less than 2000. We also apply a logarithmically-increased grid and do not find any significant difference as well.  

When the ejecta reaches the age of several tens of days there are favourable conditions for dust formation in a gas with $n\sim 10^8-10^{10}$~cm$^{-3}$, $T \simlt 10^4$~K \citep[e.g.][and many others]{Kozasa1989,Todini2001,Nozawa2006}. The dust formation continues up to $\sim 10^3$~d \citep{Todini2001}. At this age we inject the dust particles in the ejecta. We distribute the particles in the ejecta core along the radius uniformly. We add dust grains in the shells inside the core radius $r_c$ only. Initially we assume the velocity of particles is equal to that of the surrounding gas. Thus, the grains move with the free expanding ejecta. Due to the interaction with gas they support their velocity equal to that of the surrounding gas flow, that is controlled by the stopping time scale \citep{Epstein1924,Baines1965}. We include the drag force acting on grains \citep{Draine1979b}, but, for simplicity, we do not include the backward reaction on the gas. We take into account the destruction of dust particles by both thermal (in a hot gas) and kinetic (due to a relative motion between gas and grains) sputtering \citep{Draine1979a}. We consider the dynamics of grains  with size range $0.01-1~\mu$m in $N_s = 11$ binned equally in the log-scale.

The dynamics of the grains is governed by  the system of ODEs:
\ba
 {d\pmb{x}_p \over dt} = \pmb{v}_p \\
 {d\pmb{v}_p \over dt} = {\pmb{v}_{gas} - \pmb{v}_p \over \tau_s}
\label{dust-evol}
\ea
where $\pmb{x}_p$ and $\pmb{v}_p$ are the dust particle position and velocity, the stopping time is written for the Epstein drag with the supersonic Baines correction \citep{Epstein1924,Baines1965,Draine1979b}:
\be
 \tau_s = {m_p \over \pi a^2 \rho_{gas} \xi |v_p - v_{gas}|},
\label{eq-tstop}
\ee
where $m_p$, $a_p$ and $v_p$ are the mass, size and velocity of a dust particle, respectively, $\rho_{gas}$ and $v_{gas}$ are the density and velocity of gas, and the correction is 
\be
 \xi = \left[1 + {128 k_B T_{gas} \over 9 \pi m_H (v_p - v_{gas})^2}\right]^{1/2}
\ee
where $k_B$ is the Boltzmann constant, $T_{gas}$ is the temperature of gas, $m_H$ is the mass of the proton. The equations (\ref{dust-evol}) are solved self-consistently with the 1D gas dynamics. Here we study the dynamics of dust particles only, thus, they are considered as test particles. The evolution of the dust mass formed in the SN remnant will be considered elsewhere.

%----------------------- Section 3 -------------------------------
\section{RESULTS}\label{sec:res}

%%%%%%%%%%%%%%%%%%%%%%%%%%%%%%%%%%%%%%%%%%%%%%%%%%%%%%%%%%%%%%%%%%%%%
\begin{figure*}
\center
\includegraphics[width=15.5cm]{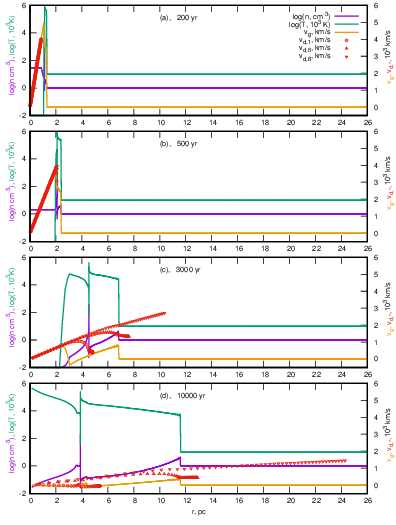}
\caption{
The radial profiles of gas density, temperature (left axis) and velocities (right axis) of gas (line) and dust grains (symbols) at $t=200$, 500, 3000 and $10^4$~yr (panels from top to bottom) for SN remnant expanding in the ambient gas with density 1~cm$^{-3}$. The symbols depict the velocity of grains with the initial size 0.01, 0.1 and 0.4~$\mu$m (circles, up and down triangles, respectively). The ejecta mass is $M_{e} = 3~\msun$.
}
\label{fig-pro-ev}
\end{figure*}
%%%%%%%%%%%%%%%%%%%%%%%%%%%%%%%%%%%%%%%%%%%%%%%%%%%%%%%%%%%%%%%%%%%%%

Figure~\ref{fig-pro-ev} presents the radial profiles of gas density, temperature and velocities of gas and dust grains at several times for SN remnant expanding in the ambient gas with density 1~cm$^{-3}$. Until $t\simlt 200$~yr the ejecta expands freely: both gas density and velocity radial profiles follow to the self-similar solution (panel (a) Figure~\ref{fig-pro-ev}). The gas density in the ejecta core has a flat distribution and  its value decreases from almost $10^{11}$~cm$^{-3}$ at 100~d to about $30$~cm$^{-3}$ at 200~yr. The gas temperature in the ejecta drops to the minimum value about 10~K. The velocity of the gas in the ejecta demonstrates a monotonic increase with distance and reaches maximum equal to $10^4$~km/s at the border between the ejecta and swept-up interstellar gas. This border is a contact discontinuity. Ahead the ejecta there is a thin shell of the swept-up interstellar gas. This shocked gas is heated up to several of $10^8$~K. Its mass is still small compared to the ejecta one. The shell is separated from the unperturbed ambient gas by the forward shock, where the jump corresponds to the Hugoniot conditions quite exactly. One can see that the ejecta has the structure typical for the free expansion phase \citep{Truelove1999}. The velocity of grains follows exactly the velocity of gas. Grains remain inside the ejecta core. The dynamics of grains is independent on their initial size.

At $t\sim 500$~yr the structure of the SN remnant has changed significantly (panel (b) of Figure~\ref{fig-pro-ev}). The core (flat part) and envelope (power-law part) of the ejecta have became indistinguishable, the gas density in these parts of the ejecta decreases down to $\sim 3$~cm$^{-3}$. The important change is the formation of the reverse shock, which  starts to move from the contact discontinuity inward the cold ejecta. The ambient gas continues sweeping up, as a result the shell becomes thicker. Another essential modification is that dust particles initially located in the core only distribute over the whole ejecta. Moreover, they overcome the contact discontinuity and move into the hot swept-up gas. 

At $t\sim 3\times 10^3$~yr these changes become more remarkable (panel (c) of Figure~\ref{fig-pro-ev}): the reverse shock heats about a half (by radius) of the ejecta, the swept-up shell gets thicker. However, the most significant modification concerns dust grains. At first, they overcome not only contact discontinuity, but also forward shock. Grains being initially larger than 0.1~$\mu$m and located in the outer parts of the ejecta core (depicted by up triangle symbols) have crossed the forward shock and reached distances up to 10~pc from the SN origin. Their velocity is ranged in $\sim 1.5-3\times 10^3$~km/s. Grains being initially smaller than 0.03~$\mu$m remain locked inside the SN remnant. Those located initially in the outer parts of the ejecta core overcome the contact discontinuity and penetrate into the swept-up shell. In the hot gas of the SN remnant grains are sputtered, especially this is efficient for small grains.

At $t\sim 10^4$~yr (panel (d) Figure~\ref{fig-pro-ev}) the structure of the SN remnant turns a common type \citep[e.g.,][]{Cioffi1988}. The ejecta is completely shocked. The gas behind the forward shock is heated up to $10^7$~K. This shock moves with velocity $\sim 300$~km/s. One should pay attention that the grains crossed the forward shock are decelerated down to $\sim 800-1500$~km/s, but this value is greater the forward shock velocity by several times. Thus, these dust particles move away further at later time. 

%%%%%%%%%%%%%%%%%%%%%%%%%%%%%%%%%%%%%%%%%%%%%%%%%%%%%%%%%%%%%%%%%%%%%
\begin{figure*}
\center
\includegraphics[width=8cm]{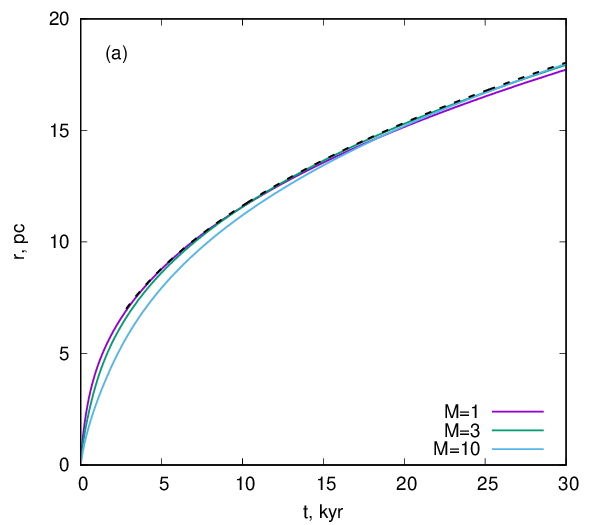}
\includegraphics[width=8cm]{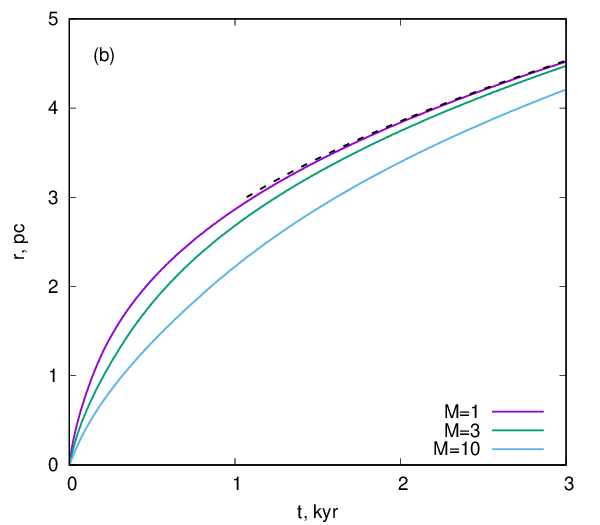}
\caption{
The radius of the remnant's shell for the SN explosion in the ambient medium with $n = 1$ and 10~cm$^{-3}$ (left and right panels, respectively) for the SN ejecta mass of $M_{e} = 1, \ 3$, and 10~$\msun$. The dash lines correspond the Sedov solution $r = (Et^2/\rho)^{1/5}$ with $E=10^{51}$~erg, $n=\rho/m_p = 1$ and 10~cm$^{-3}$. They start at time, when the swept-up mass in the SN shell is equal to $\sim 30~\msun$. 
}
\label{fig-r-evol}
\end{figure*}
%%%%%%%%%%%%%%%%%%%%%%%%%%%%%%%%%%%%%%%%%%%%%%%%%%%%%%%%%%%%%%%%%%%%%

Before we consider which of these grains will survive against sputtering and how far they run away, we put our attention on the dynamics of the SN remnant. After a short period of free expansion the mass of swept-up ambient gas in the SN shell reaches close to that of the ejecta, so that the SN remnant goes to the intermediate phase \citep{Truelove1999}. Depending on the density of the ambient gas the duration of this phase is changed, it is longer in denser medium. After the swept-up mass becomes much greater than that of the ejecta the expansion of the SN remnant follows to the self-similar adiabatic Sedov solution. The remnant remains adiabatic until radiative losses becoming significant in the shell, i.e the age gets compared to the cooling time: $t_c \sim 3\times 10^4 n^{-1}$~yr. Here we constrain our runs by this timescale. 

Figure~\ref{fig-r-evol} presents the radius of the SN remnant for several SN ejecta mass values. For low-mass ejecta of $M_{e} = 1~\msun$ the mass of the SN shell becomes about a factor of 30 more massive than the ejecta at $t\sim 3$~kyr, when the forward shock reaches about 7~pc from the SN origin (left panel). One can note that since this age the radius of the shell follows closely the radius $r = (Et^2/\rho)^{1/5}$ (see dash line at left panel). For more massive ejecta this occurs later, e.g. for $M_{e} = 10~\msun$ it takes place at SN age $\sim 10$~kyr. If the SN with massive ejecta evolves in denser medium the difference from the self-similar solution remains significant till the onset of the radiative phase (right panel, compare color lines with dash one).

%%%%%%%%%%%%%%%%%%%%%%%%%%%%%%%%%%%%%%%%%%%%%%%%%%%%%%%%%%%%%%%%%%%%%
\begin{figure*}
\center
\includegraphics[width=16cm]{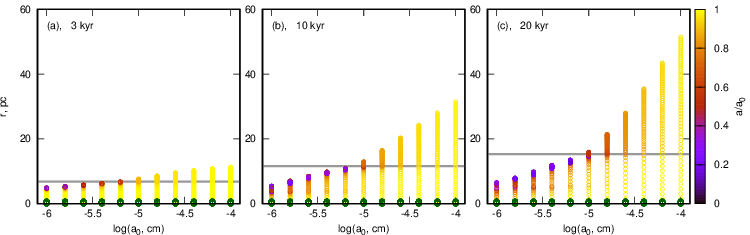}
\caption{
The radial distance (from the SN origin), which grains with the initial size of $a_0$ reach at the age 3, 10 and 20~kyr (panels from left to right) of the SN with the ejecta mass $M_{e} = 3~\msun$. The radial distances at early time ($t=200$~yr) are shown by green symbols (see near the $x$-axis). Color bar shows the ratio of the current size of a grain to its initial value. The horizontal line depicts the current SN shell radius. The ambient gas density is 1~cm$^{-3}$.
}
\label{fig-rd-size-m3-evol}
\end{figure*}
%%%%%%%%%%%%%%%%%%%%%%%%%%%%%%%%%%%%%%%%%%%%%%%%%%%%%%%%%%%%%%%%%%%%%

In Figure~\ref{fig-pro-ev} one can find only positions for grains of three sizes. During their moving through the swept-up SN shell grains can be sputtered. Therefore, we need to consider not only dynamics, but also destruction of grains in details. Figure~\ref{fig-rd-size-m3-evol} presents how far grains run away from the SN origin and how efficiently they can be destroyed during their movement. At $y$-axis there is the radial distance from the SN origin, which grains with the initial size of $a_0$ (the ${\rm log}~a_0$ is at $x$-axis) reach at different age of the SN remnant. Using the color bar, which shows the ratio of the current size of a grain to its initial value, one can estimate the destruction efficiency. At the SN age early as $200$~yr grains are located inside the cold core of the SN ejecta (see also panel (a) in Figure~\ref{fig-pro-ev}), in Figure~\ref{fig-rd-size-m3-evol}a their radial positions are depicted by green points. 

At the SN age of 3~kyr dust grains smaller than $0.03~\mu$m are located inside the SN remnant: their radial distances are smaller than the SN radius (shown by grey line). Those grains being inside the swept-up shell (at $r \sim 5$~pc, see also panel (c) in Figure~\ref{fig-pro-ev}) are sputtered in the hot gas, their sizes decrease about a factor of $2-3$. While the grains of $a_0\simlt 0.03~\mu$m located initially in the inner parts of the ejecta do not still cross the contact discontinuity (which has the radius $\sim 4.5$~pc, see panel (c) in Figure~\ref{fig-pro-ev}). The size of these grains remains unchanged or very close to the initial one. One should note that grains larger than $a_0 \sim 0.05~\mu$m overcome the forward shock front and get into the unperturbed ambient gas, where they will be not subjected to thermal sputtering. While the kinetic sputtering can influence on these grains due to they have high velocities relative to the ambient gas at rest. However, its timescale exceeds several kyrs for $a_0 \sim 0.1~\mu$m and several tens kyrs for $a_0 \sim 1~\mu$m. Thus, at early SN ages grains with $a_0 \simgt 0.1~\mu$m survive againt sputtering. 

\begin{figure*}
\center
\includegraphics[width=16cm]{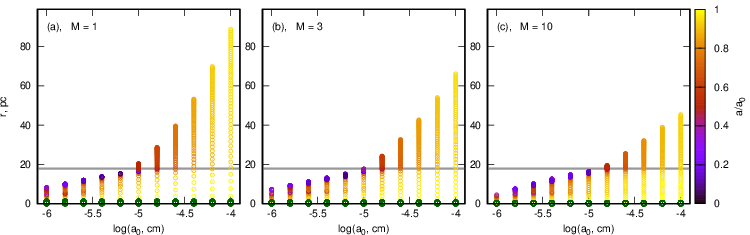}
\caption{
The radial distance (from the SN origin), which grains with the initial size of $a_0$ reach at SN age $30$~kyr for the SN ejecta mass $M_{e} = 1, \ 3$, and 10~$\msun$ (panels from left to right). The radial distances at early time ($t=200$~yr) are shown by green symbols (see near the $x$-axis). Color bar shows the ratio of the current size of a grain to its initial value. The horizontal line depicts the current SN shell radius. The ambient gas density is 1~cm$^{-3}$.
}
\label{fig-rd-size1}
\end{figure*}

At older SN remnant grains initially small as $a_0\simlt 0.03~\mu$m are in hot gas behind the reverse shock, where they are sputtered. Those grains, which cross the contact discontinuity, are efficiently destroyed in hostile conditions of the hot SN shell. A part of these grains is completely sputtered there. Larger grains with initial size of $a_0\sim 0.1~\mu$m survive behind the forward shock, moreover their size decreases slightly. Only grains close to the forward front and crossing it are destroyed significantly. At $t\sim 20$~kyr their size decreases a factor of $1.5-2$ compared to the initial value (Figure~\ref{fig-rd-size-m3-evol}c) and becomes smaller about $3-5$ times at $30$~kyr (Figure~\ref{fig-rd-size1}b). These grains reach radial distances larger about a factor of $\sim 1-1.5$ than the SN forward shock radius. 

Grains with size of $a_0\sim 1~\mu$m save their sizes almost unchanged during the whole evolution considered here (see Figure~\ref{fig-rd-size-m3-evol} and the middle panel of Figure~\ref{fig-rd-size1}). Their high velocity allow them to reach radial distances exceeding two times the SN shell radius at $t=3$~kyr. This exceeding factor increases during the evolution: it reaches $\sim 3$ at $20$~kyr (Figure~\ref{fig-rd-size-m3-evol}c) and more than $\sim 4$ at $30$~kyr (the middle panel of Figure~\ref{fig-rd-size1}).

%%%%%%%%%%%%%%%%%%%%%%%%%%%%%%%%%%%%%%%%%%%%%%%%%%%%%%%%%%%%%%%%%%%%%
%%%%%%%%%%%%%%%%%%%%%%%%%%%%%%%%%%%%%%%%%%%%%%%%%%%%%%%%%%%%%%%%%%%%%

%%%%%%%%%%%%%%%%%%%%%%%%%%%%%%%%%%%%%%%%%%%%%%%%%%%%%%%%%%%%%%%%%%%%%
\begin{figure*}
\center
\includegraphics[width=16cm]{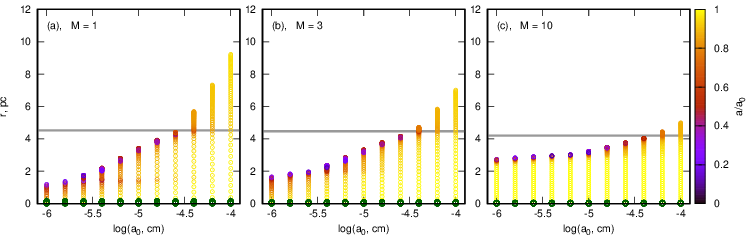}
\caption{
The same as in Figure~\ref{fig-rd-size1}, but for the ambient gas density equal to 10~cm$^{-3}$ and the SN remnant of age $t=3$~kyr.
}

Эволюция массы пыли, образованной в остатке СН, будет рассмотрена отдельно
\label{fig-rd-size10}
\end{figure*}
%%%%%%%%%%%%%%%%%%%%%%%%%%%%%%%%%%%%%%%%%%%%%%%%%%%%%%%%%%%%%%%%%%%%%

For another value of the ejecta mass the dynamics of grains is qualitatively similar to that presented in Figure~\ref{fig-rd-size-m3-evol} for $M_e = 3~\msun$. The maximum velocity of gas in the SN ejecta is higher for less massive ejecta (Figure~\ref{fig-ini-pro}). Therefore, grain velocities at the SN free-expansion phase increases as $M_e^{-0.5}$. This does not significantly influence on grains initially smaller than $a_0\sim 0.1~\mu$m, because they spend their lifetime mostly behind the forward shock. They are desroyed efficiently during their movement through the hot SN shell. Only grains with $a_0\sim 0.04-0.06~\mu$m can reach closely to the forward shock front, but they cannot cross it (Figure~\ref{fig-rd-size1}). As it is mentioned, grains being initially larger than $a_0\sim 0.1~\mu$m overcome the shock front. For less-massive ejecta such grains can run away on larger distance, which is greater for larger size of grain. For example, grains with size of $a_0\sim 1~\mu$m can reach $\sim 90$~pc for $M_e=1~\msun$ (panel a), while this distance decreases twice for $M_e=10~\msun$ (panel c).

During the SN expansion in denser ambient gas, grains with initial size of $a_0\simgt 0.1~\mu$m can overcome the forward shock, but they do not spread on large distances (Figure~\ref{fig-rd-size1}). For the SN with $M_e =1~\msun$ expanding in the ambient medium with $n=10$~cm$^{-3}$ grains with $a_0\sim 1~\mu$m can reach about $\sim 9$~pc, this is only two times larger than the SN shell radius (Figure~\ref{fig-rd-size10}a). While such grains in less dense environment with $n=1$~cm$^{-3}$ run away to distance about four times larger than the shell radius (Figure~\ref{fig-rd-size1}a). For the ejecta massive as $10~\msun$ such grains attain distances about 20\% larger than the shell radius (Figure~\ref{fig-rd-size10}c).

%%%%%%%%%%%%%%%%%%%%%%%%%%%%%%%%%%%%%%%%%%%%%%%%%%%%%%%%%%%%%%%%%%%%%
\begin{figure*}
\center
\includegraphics[width=16cm]{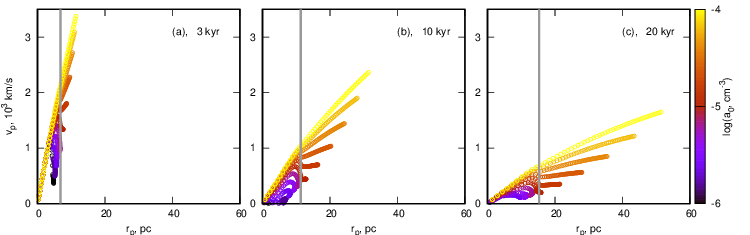}
\caption{
The 'radius--velocity' diagram for dust particles with the initial size of $a_0$ at the SN age 3, 10 and $20$~kyr (panels from left to right) for the SN ejecta mass $M_{e} = 3~\msun$. Color bar shows the initial size of a grain. The vertical line depicts the current SN shell radius. The ambient gas density is 1~cm$^{-3}$. 
}
\label{fig-rd-vd-m3-evol}
\end{figure*}
%%%%%%%%%%%%%%%%%%%%%%%%%%%%%%%%%%%%%%%%%%%%%%%%%%%%%%%%%%%%%%%%%%%%%

%%%%%%%%%%%%%%%%%%%%%%%%%%%%%%%%%%%%%%%%%%%%%%%%%%%%%%%%%%%%%%%%%%%%%
\begin{figure*}
\center
\includegraphics[width=16cm]{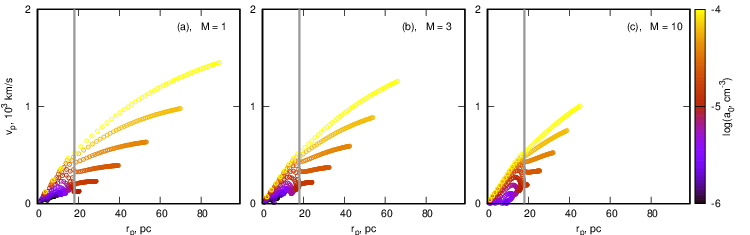}
\caption{
The 'radius--velocity' diagram for dust particles with the initial size of $a_0$ at the SN age $30$~kyr. Color bar shows the initial size of a grain. The panels from left to right correspond to the SN ejecta mass $M_{e} = 1, \ 3$, and 10~$\msun$. The vertical line depicts the current SN shell radius. The ambient gas density is 1~cm$^{-3}$.
}
\label{fig-rd-vd1}
\end{figure*}
%%%%%%%%%%%%%%%%%%%%%%%%%%%%%%%%%%%%%%%%%%%%%%%%%%%%%%%%%%%%%%%%%%%%%

%%%%%%%%%%%%%%%%%%%%%%%%%%%%%%%%%%%%%%%%%%%%%%%%%%%%%%%%%%%%%%%%%%%%%
\begin{figure*}
\center
\includegraphics[width=16cm]{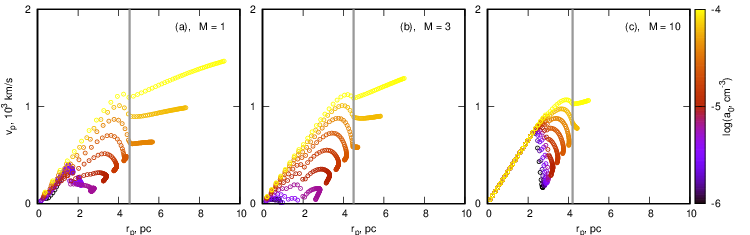}
\caption{
The same as in Figure~\ref{fig-rd-vd1}, but for the ambient gas density equal to 10~cm$^{-3}$ and the SN remnant of age $t=3$~kyr.
}
\label{fig-rd-vd10}
\end{figure*}
%%%%%%%%%%%%%%%%%%%%%%%%%%%%%%%%%%%%%%%%%%%%%%%%%%%%%%%%%%%%%%%%%%%%%

Grains get their initial velocities from a gas in which they are formed at the SN free-expansion phase. We assume that grains are formed inside the ejecta core, so the maximum velocity is $v_{d,0} \sim v_e (r_c/R_e)$ and the velocity increases to this value linearly with radius (see eq.~\ref{eq-vel}). Since $t\sim 1000$~days, when grains are supposed to be formed, grains remain tightly coupled with surrounding gas. This condition is violated when the gas profile begins to be modified by the interaction with the ambient gas, i.e. a significant mass of such gas is swept up in the SN shell (Figure~\ref{fig-pro-ev}b). After this moment grains of different size start to evolve separately. Small grains remain coupled, they respond quickly on changes in surrounding gas due to the stopping time depends on grain size as $\tau_s \sim a$ (eq.~\ref{eq-tstop}). But small grains spend their lifetime inside the SN remnant, where they are significantly destroyed and their relation "radius--velocity" shows the linear dependence (see magenta symbols in Figure~\ref{fig-rd-vd-m3-evol}).

Large grains continue to move freely, that allows them to cross the contact discontinuity, pass through the hot swept-up shell and overcome the SN forward shock. Therefore, they can move in the ambient unperturbed gas. During this process the velocity of grains with size of $a_0 \sim 0.1~\mu$m decreases, but remains high enough (see red symbols in Figure~\ref{fig-rd-vd-m3-evol}). While grains large as $a_0 \sim 1~\mu$m demonstrate linear relationship between their velocity and radial position (see yellow symbols in Figure~\ref{fig-rd-vd-m3-evol}). Similar relation remains at the age $30$~kyr irrespective of the ejecta mass (cf. panels in Figure~\ref{fig-rd-vd1}). At such late age the dependence on grain size becomes more clear. Increase of the ambient density leads to break the linear relation at distance close to the SN shell radius (cf. panels in Figure~\ref{fig-rd-vd10}. For grain with initial size of $a_0 \sim 1~\mu$m the relation remains linear inside and outside the SN radius, but it has different slopes.

%%%%%%%%%%%%%%%%%%%%%%%%%%%%%%%%%%%%%%%%%%%%%%%%%%%%%%%%%%%%%%%%%%%%%
\begin{figure*}
\center
\includegraphics[width=8cm]{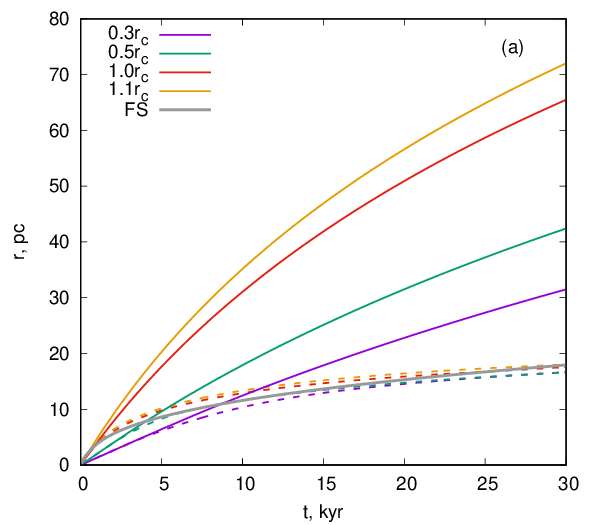}
\includegraphics[width=8cm]{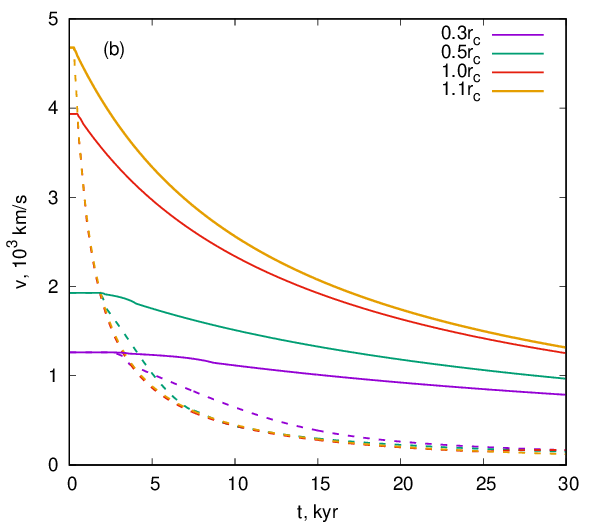}
\caption{
The radial position and velocity of a grain initially located at distance $0.3,\ 0.5,\ 1.0$ and $1.1~r_c$ (shown by lines of different color). The tracks for a grain with initial size $a_0$ equal to 0.1 and 1~$\mu$m are depicted by dashed and solid lines, respectively. The grey line shows the SNR radius, i.e. the position of the forward shock.
}
\label{fig-r-v-track}
\end{figure*}
%%%%%%%%%%%%%%%%%%%%%%%%%%%%%%%%%%%%%%%%%%%%%%%%%%%%%%%%%%%%%%%%%%%%%

Grains born at various radial distance got distinct initial velocities (see Figure~\ref{fig-pro-ev}a and~\ref{fig-ini-pro}). One can found the differences in Figure~\ref{fig-rd-vd1}. However, for clarity, 
Figure~\ref{fig-r-v-track} presents the evolution of the position and velocity of a grain initially located at various radial distances. Grains of size $0.1~\mu$m born at the edge of the ejecta core ($\sim 1~r_c$) remain inside the SN remnant until the age $\sim 1$~kyr (see red dash line in panel (a) of Figure~\ref{fig-r-v-track}).  Afterwards, they cross the forward shock front and move in the ambient interstellar gas. Grains born in the interiors of the ejecta core at $\sim (0.3-0.5)~r_c$ pass through the front at $\sim 6-16$~kyr (green and magenta lines in panel (a) of Figure~\ref{fig-r-v-track}). After they leave the SN remnant, they move with almost the same velocity about $\sim 400$~km/s (see dash lines). Note the flat parts of the velocity profiles correspond to dynamical coupling grains with gas (panel (b) of Figure~\ref{fig-r-v-track}). Grains of size $1~\mu$m move with larger velocity due to they are decoupled from the gas earlier (solid lines in panel (a) of Figure~\ref{fig-r-v-track}). Their velocities mainly exceed 1000 km/s and tend also to a certain value, but it will be reached much later than 30~kyr (panel b).

Above we assume that grains are born in the ejecta core. Although gas density decreases rapidly (as $r^{-9}$) in the ejecta envelope (Figure~\ref{fig-ini-pro}), the conditions favourable for grain formation may be inside the envelope as well, especially while the gas density does not decrease remarkably. Thus, we consider the evolution of grains born in the ejecta envelope at $\sim 1.1~r_c$. At this distance the gas density decreases twice compared to the value in the core. Grains of size $0.1~\mu$m born at this distance demonstrate the same evolutionary tracks as grains formed at the core edge $\sim 1~r_c$ (compare yellow and red lines in panel (a) of Figure~\ref{fig-r-v-track}). While larger grains of size $1~\mu$m spread over greater distances due to keeping higher velocity (see yellow solid lines). Thus, one can conclude that larger grains born close to the border between core and envelope of the SN ejecta leave the SN remnant with the highest velocity.

\begin{figure*}
\center
\includegraphics[width=16cm]{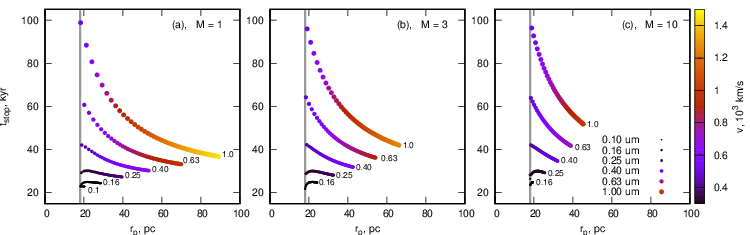}
\caption{
The 'radius--stopping time' diagram for dust particles with the initial size of $a_0$ (depicted by symbols of different size and numbers near the symbols) located outside the SN remnant of age $30$~kyr. Color bar shows the current velocity of a grain. The panels from left to right correspond to the SN ejecta mass $M_{e} = 1, \ 3$, and 10~$\msun$. The vertical line depicts the current SN shell radius. The ambient gas density is 1~cm$^{-3}$. 
}
\label{fig-rd-tstop1}
\end{figure*}

\begin{figure*}
\center
\includegraphics[width=16cm]{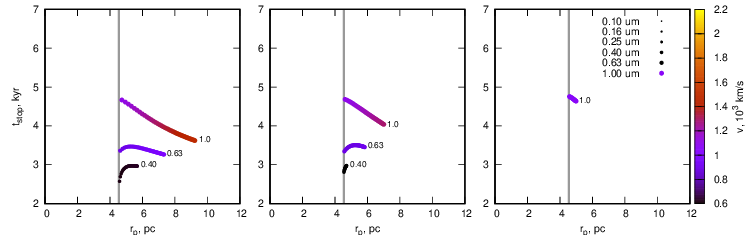}
\caption{
The same as in Figure~\ref{fig-rd-tstop1}, but for the ambient gas density is 10~cm$^{-3}$ and the SN remnant of age $t=3$~kyr.
}
\label{fig-rd-tstop10}
\end{figure*}

Large grains crossing the forward shock have velocities from several hundred to $1-2$ thousand km/s (Figures~\ref{fig-rd-vd1}-\ref{fig-rd-vd10}). Such high values allow them to move at large distances from the SN remnant. These grains keep their high velocity during long period. Using eq.~\ref{eq-tstop} we estimate the stopping time for grains passed from the remnant to the ambient medium. Figure~\ref{fig-rd-tstop1} presents the stopping time for grains, which overcome the forward shock. For the ambient gas density of 1~cm$^{-3}$ grains with size $a_0 \sim 0.1~\mu$m (the smallest symbols in Figure~\ref{fig-rd-tstop1}) are located at distances $\sim 18-30$~pc at the SNR age $30$~kyr. These grains decelerate within stopping time $\sim 50-60$~kyr. During this period the grains with velocity  $\sim 300$~km/s can propagate up to 40~pc. Grains large as $a_0 \sim 1~\mu$m have already passed longer distance till $30$~kyr, they can be found at $r\sim 40-90$~pc depending on the SN ejecta mass. Their velocity keeps higher than $10^3$~km/s, therefore during stopping time of $\sim 40-60$~kyr such large grains can reach distance up to 100~pc and even exceed it. Such distance is about a factor of several greater than the SNR radius. In more dense medium grains spread out smaller distances, which only exceeds the remnant radius a factor of two or less (Figure~\ref{fig-rd-tstop10}).

%%%%%%%%%%%%%%%%%%%%%%%%%%%%%%%%%%%%%%%%%%%%%%%%%%%%%%%%%%%%%%%%%%%%%
%%%%%%%%%%%%%%%%%%%%%%%%%%%%%%%%%%%%%%%%%%%%%%%%%%%%%%%%%%%%%%%%%%%%%

%%%%%%%%%%%%%%%%%%%%%%%%%%%%%%%%%%%%%%%%%%%%%%%%%%%%%%%%%%%%%%%%%%%%%
%%%%%%%%%%%%%%%%%%%%%%%%%%%%%%%%%%%%%%%%%%%%%%%%%%%%%%%%%%%%%%%%%%%%%

%----------------------- Section 4 -------------------------------
\section{DISCUSSION}\label{sec:dis}

In a homogeneous unmagnetized medium grains left the SNR move radially outside from the SN origin. However, in more realistic conditions there are processes, which prevent grains to keep high velocity and propagate at large distances. Grains can be decelerated in dense gas of circumstellar structures like extended wind shell and dense clumps. In any case the motion of grains in this environment depends on filling factor of dense gas. Early interaction of massive stars with ambient gas by high ionizing radiation flux and strong wind has mostly cleaned up from dense clumps of the parent molecular cloud. However, dense circumstellar wind might be sufficiently extended for decelerating grains of $\simlt 1~\mu$m, while larger ones can pass through the wind and move further. 

More substantial obstacle for grains is turbulent magnetic fields. Grains larger $0.1~\mu$m, which are able to leave the SNR, are charged \citep{Weingartner2001} and, consequently, tightly coupled with magnetic field. This can lead to their acceleration \citep{Lazarian2002,Yan2004} or deceleration \citep{McKee1987}. Charged grains go through many reflections owing to interaction with magnetic structures, some of them can move backward to the SNR and demonstrate “pinball” motion \citep{Fry2020}, which looks like a random walk. Although we should nota that for large dust grains escaping the SN remnant the effect of charge on their dynamics will be suppressed, since the charge-to-mass ratio of the grain decreases with its size: $q/m \sim a^{-2}$. For example, large grains $\simgt 1~\mu$m can penetrate inside the heliosphere \citep[e.g.,][]{Slavin2012,Sterken2022,Godenko2023}. The efficiency of these processes obviously depends on the structure of magnetic field. This question requires more detailed study similar to the issue related to grain destruction by shattering. On one hand, grains can efficiently collide each other and shatter into more smaller ones \citep{Jones1996,Hirashita2009,Murga2019}, but in turbulent magnetized flows an enhanced gas--grain coupling from the Lorentz force can provide to grains relative protection from shattering \citep{Moseley2023}. 

Cosmic rays and ultraviolet radiation can change the charge of dust grains \citep{Weingartner2001,Shchekinov2007,Ivlev2015}. While cosmic rays are significant only in fairly dense regions of molecular clouds with density of $10^4-10^6$~cm$^{-3}$, which is remarkably higher than the values in the SN shell Although cosmic rays are accelerated in the remnant shell, but large dust grains spend a relatively short time ($\sim \Delta r / v_d \simlt 10^3$~yr) compared to the average time for interaction with cosmic rays ($\simgt 10^5$~yr). The acceleration is facilitated by the complex structure of magnetic fields in young SNe remnants, and the field strength varies significantly according to the polarimetric observations of remnants in the Galaxy \citep[see for review][]{Slane2024}. The ultraviolet radiation field, in turn, also experiences noticeable fluctuations in the interstellar medium due to the non-uniform distribution of its sources, i.e. massive stars, and absorption systems, namely, molecular clouds and dust. A comprehensive study of the mutual influence of the above-mentioned processes requires a three-dimensional approach that takes into account the transport of dust, radiation, and cosmic rays, which is clearly out of the scope of the simple approach presented here.

High-velocity grains moving vertically out from the galactic midplane are able to reach the galactic halo and go out from the parent galaxy to the circum/intergalactic medium (CGM/IGM). Thus, such grains from SN can be one more source for the IGM dust among others \citep[e.g.,][]{Shchekinov2011}. Their transport is provided through their high velocity got at the birthplace in the SN ejecta. During their motion in the ISM they are not only decelerated, but may obtain extra portion of momentum due to radiation pressure, which obviously dominates in dust ejection from galaxies \citep{Aguirre2001,Ferrara1990,Ferrara1994,Shustov1995,Sharma2011,Hirashita2019}. Large high-velocity grains can hardly affect on the IGM extinction \citep{Inoue2003}, however, they may be destroyed by sputtering in the CGM and afterwards being already smaller they reach the IGM.

%----------------------- Section 5 -------------------------------
\section{CONCLUSIONS}\label{sec:con}

Here we have considered the dynamics of grains formed in the SN ejecta from their birth to the age of several thousand years, i.e. when the SN shell becomes radiative, using a one-dimensional spherically symmetric Lagrangian gas dynamics coupled with dust particle dynamics. The results are summarized as follows:
\begin{itemize}
 \item dust grains smaller than $0.03~\mu$m born in the SN ejecta core penetrate in the ejecta envelope and into the swept-up shell, afterwards, these grains remain locked inside the SN remnant, where they are  efficiently destroyed mainly by thermal sputtering inside hot SN remnant gas heated by both forward and reverse shocks;
 \item grains having the initial size equal to $\sim 0.1~\mu$m can overcome the forward shock and spread over the unperturbed ISM with velocity about several hundred km/s for the SN expanding in an ambient gas with density of $1$~cm$^{-3}$; during their moving through the hot SN shell these grains decelerate efficiently and their size decreases in several times due to sputtering;
%%%% \item grains located initially in the inner parts of the ejecta have not crossed the contact discontinuity at least while the SN remnant remains non-radiative, e.g. at the age of 30~kyr for the SN expanding in an ambient gas with density of $1$~cm$^{-3}$; although such grains are inside hot bath behind the reverse shock, their size remains unchanged or very close to the initial one;
 \item grains large as $\simgt 0.3~\mu$m can overcome the forward shock and spread over the unperturbed ISM, they retain their size almost constant independently on where such grains locate at their formation inside the ejecta core; 
 \item the velocity of such large grains vary from several hundred to $1-2$ thousand km/s, these grains pass through the SN shell without significant deceleration and continue to move at high velocity in the ambient ISM, e.g. for the ambient gas density of 1~cm$^{-3}$ grains large as $a_0 \sim 1~\mu$m can be found at distances from the SN origin $r\sim 40-90$~pc at the age of $\sim 30$~kyr, this distance is about a factor of $2-5$ longer than the current SN shell radius ($\sim 17.5$~pc); for velocity higher than $10^3$~km/s their stopping time is $\sim 40-60$~kyr, therefore, during further evolution these grains can reach distance up to 100~pc from the SN origin and even exceed it;
 \item in dense environment grains spread over significantly smaller volume around the SNR, e.g. for the SN with the ejecta of $3~\msun$ evolving into the ambient gas density of 10~cm$^{-3}$ grains with $a_0 \simgt 0.3~\mu$m can overcome the forward shock, but they remain close to the shock front, while even larger grains with $a_0 \sim 1~\mu$m reach distances only a factor of $1.5$ greater than the current SN shell radius ($\sim 4.5$~pc);
 \item the propagation distance for large grains depends on the mass of the SN ejecta: in less-massive ejecta the velocity of a gas, in which grains are formed at the SN free-expansion phase, is supposed to be higher, therefore, these grains have got higher initial velocity and can spread over greater volume of the ISM;
 \item  grains larger $1~\mu$m born close to the outer border of the SN ejecta core leave the SN remnant with the highest velocity.
\end{itemize}
Thus, high-velocity grains are expected to be formed on the border between core and envelope of low-mass SN ejecta with mass $\simlt 3~\msun$. If this SN evolves in a diffuse environment with density low as $\simlt 1$~cm$^{-3}$, grains with initial size of $\sim 1~\mu$m can spread over distances exceed 100~pc from the SN origin at timescale of $\simgt 30$~kyr after SN explosion and their velocity keeps higher than several hundreds km/s, their sizes do not decrease significantly during this path. High-velocity grains with much larger size are expected to travel across the ISM several hundreds kiloyears or maybe even longer. Therefore, the meteor detected on the 6-meter telescope \citep{Afanasiev2007} might belong to this population of the interstellar grains. Note that a grain moving with about one thousand km/s can get away up to 1~kpc from its birthplace within one million years. Thus, a part of such grains can leave the Milky Way disk and reach the halo. 

%----------------------- Section A -------------------------------
\section*{Acknowledgements}

The author warmly remembers long discussions with Victor Afanasiev on a variety of questions related to possible origin of high-velocity dust grains. The author thanks to the reviewers for their careful reading and important comments. The author is grateful to Yuri Shchekinov for many discussions and long-term support, Svyatoslav Dedikov for discussions and Ilya Khrykin for his assistance.

%\section*{FUNDING}

\section*{CONFLICT OF INTEREST}

The authors of this work declare that they have no conflicts interest.

%----------------------- Section L -------------------------------
\bibliographystyle{aspb1}
\bibliography{p-bib1.bib}

\end{document}